\documentclass[referee]{raa}            % referee version: for submission

\usepackage{graphicx,times}             %for PS/EPS graphics inclusion, new
\usepackage{color}
\usepackage{natbib}
\usepackage{amssymb,amsmath}
\usepackage{enumerate}				%bianhao
\usepackage{indentfirst}			%suojing

\bibpunct{(}{)}{;}{a}{}{,}

\usepackage[a4paper=true,pagebackref=true,driverfallback=dvipdfm]{hyperref}

\begin{document}

   \title{Optimal subreflector position determination of shaped dual-reflector antennas based on the parameters iteration approach
%\footnotetext{$*$ Supported by the National Natural Science Foundation of China.}
}
%   \subtitle{I. Place Your Subtitle Here}

   \volnopage{Vol.0 (20xx) No.0, 000--000}      %%preserved for Editor. DOn't remove!
   \setcounter{page}{1}          %%starting page, preserved for Editor. DOn't remove!

   \author{Bin-bin Xiang
      \inst{1,2}
   \and Cong-si Wang
      \inst{1}
   \and Pei-yuan Lian
      \inst{1}
   \and Na Wang
      \inst{2}
    \and You Ban
      \inst{1}
   }

   \institute{Key Laboratory of Electronic Equipment Structure Design of Ministry of Education, School of Mechano-Electronic Engineering, Xidian University, Xi’an, Shaanxi 710071, China; {\it congsiwang@163.com}\\
        \and
    Xinjiang Astronomical Observatory, Chinese Academy of Sciences, Urumqi, Xinjiang 830011, China\\
\vs\no
   {\small Received~~20xx month day; accepted~~20xx~~month day}}

\abstract{A new method based on parameters iteration technique has been developed to determine the optimal subreflector position for shaped Cassegrain antennas to improve the electromagnetic (EM) performance distorted by gravity. Both the features of shaped surface and the relationship between optical path difference (OPD) and far field beam pattern are employed. By describing the shaped dual-reflector surface as a standard discrete parabola set, we can utilize the optical features of the standard Cassegrain system in the classical OPD relationship. Then, the actual far field beam pattern is expressed as the synthesis of ideal beam and error beam by decomposing subreflector adjustment parameters using mechanical-electromagnetic-field-coupling-model (MEFCM). Furthermore, a numerical method for determining optimal subreflector position is presented. The proposed method is based on the iteration technique of subreflector adjustment parameters, and the optimal far field pattern is used as the iteration goal. The numerical solution of optimal adjustment parameters can be obtained rapidly. Results of a 25 m Shaped Cassegrain antenna demonstrate that the adjustment of the subreflector to the optimal position determined by the proposed method can improve the EM performance effectively.
}

   \authorrunning{B. B. Xiang, C. S. Wang, N. Wang, P. Y. Lian \& Y. Ban }            %author_head in even pages
   \titlerunning{Optimal subreflector position determination }  % title_head in odd pages

   \maketitle
%% The author head (on even pages) and the title head (on odd pages) will be
%% automatically extracted from \author{} and \title{}. Whenever the title is too long,
%% you will be asked to supply a shorter one by inserting either \authorrunning{} or
%% \titlerunning{} before \maketitle. Anyway, you can specify your own heads.
%%
%%
%% Note: In the following text body of your manuscript, please note several differences from
%%       other major journals:
%% (1) \subsection{Please Capitalize the First Letter of Each Notional Word in Subsection Title}
%% (2) Please Capitalize the First Letter of Each Notional Word in all tables' captions

%%%%%%%%%%%%%%%%%5=============================contect
%
%________________________________________________ sections below
%
%=======================Section 1===============================
\section{Introduction}           %% first-level sections will be auto-capitalized
\label{sect:intro}

Structural deformation of large reflector antenna is inevitable in operation conditions due to exterior loads such as gravity, temperature and wind, which will result in antenna gain loss and pointing error caused by beam distortions. The mechanical-electromagnetic-field-coupling-model (MEFCM) (\citealt{Duan+Wang+2009, Lian+etal+2015, Wang+etal+2017}) is widely applied to analyse the influence of structural de-formation on the electromagnetic (EM) performance of reflector antennas. It can be used to rapidly analyse the influence of different types of errors on far field beam pat-tern, such as surface random errors and structural deformation errors (\citealt{Wang+etal+2007}). In MEFCM, a key step is to obtain the aperture field phase error (PE) or optical path difference (OPD), especially for dual-reflector antennas (\citealt{Ban+etal+2017}).

Many researchers have studied the influence of structural errors on OPD and EM performance of reflector antennas. Duan (\citealt{Duan+Wang+2009}) studied the influ-ence of surface random errors and systematic errors on EM performance and established the optimization model for the integrated mechanical-electromagnetic performance. Baars (\citealt{Baars+2007}) studied the influence of different types of structural deformation on OPD for primary-focus antennas, such as axial and lateral feed defocus errors. Ruze (\citealt{Ruze+1969}) studied the influence of different types of structural deformation on the EM performance for dual-reflector antennas and derived the relationship between different types of structural errors and OPD for dual-reflector, such as feed displacement, sub reflector translation and rotation offsets, etc.

In order to realize high gain and favourable beam pattern for large-aperture and high-frequency reflector antennas, many researchers focus on the active compensation technique through feed or sub reflector adjustment. In 2009, Duan  (\citealt{Song+etal+2009,Zhang+etal+2017}) proposed a computational model for analysing the effect of both reflector errors and phase centre errors on far field pattern to find the optimal phase centre. A method for describing the relationship between far field and aperture field by the aperture field integration method was presented in (\citealt{Lian+etal+2014}) and the feed adjustment quantity was determined from far pattern by proposed method. In these literatures, the researches were mainly about the correction of the feed position and pose based on deformation of primary reflector surface. However, for dual reflector antennas, the feeds are difficult to be adjusted frequently because they are located on secondary focus near the primary reflector vertex, which are large and heavy. In fact, adjustment of the sub reflector position would be an easier way to improve aperture efficiency. For the problem of EM performance degradations caused by primary reflector deformation, the group of approximation paraboloids were used to fit the deformed surface in (\citealt{Wang+etal+2013}). The new sub reflector position was obtained by utilizing an optimal geometric match to compensate the primary surface deformation. In (\citealt{Doyle+Keith+2009}), sub reflector or feed was moved to compensate the effect of antenna structural-thermal deformation. The common premise in (\citealt{Wang+etal+2013}) and (\citealt{Doyle+Keith+2009}) was the deformation information of the primary reflector in most working conditions. Nevertheless, the environment of antenna work was very complicated, including gravity, thermal and wind load, and it was not easy to obtain the actual shape of antenna surface quickly. Generally, the surface deformation of the antenna can be measured by industrial photogrammetry, but the measurement needs to use crane for assistance and it only works at night, and the operation is difficult and needs a long time to complete the measurement, so the deformed surface shape within the scope of all elevation can not be obtained in a short time. Therefore, it is desirable to correct the antenna deformation by adjusting the subreflector, which is quick and simple. In this way, the EM performance of the antenna is obtained directly from the receiver and terminal of the radio telescope antenna, and the real-time subreflector position adjustment can be realized based on electromechanical coupling theory to improve the antenna efficiency.

In the electromechanical coupling theory, the relationship between antenna structural deformation and OPD should be used in MEFCM to analyse the influence of structural errors on far field beam pattern. However, these relationships (\citealt{Ruze+1969,Baars+2007,Duan+Wang+2009}) are only suitable for standard reflectors, which can be accurately described by a closed-form expression, and they cannot be directly applied for shaped reflectors. In this paper, the shaped surface described by discrete standard parabola set was adopted, and a new method for optimal subreflector position determination based the iteration technique was proposed, which was aimed to obtain the optimal EM performance. The optimal parameters of the subreflector position can be obtained rapidly by efficient numerical iterative algorithm and fast numerical computation solution. Numerical computation results for a 25 m shaped Cassagrain antenna indicate that the reduction of EM performance caused by gravity deformation can be substantially compensated by adjustment of subreflector position and the adjustment parameters obtained by the iteration approach is effective and appropriate.

%=======================Section 2===============================
\section{Optical path difference of shaped dual-reflector antenna}
\label{sect:Set2}

\begin{figure}[htb]
	\centering
	\includegraphics[width=8cm, angle=0]{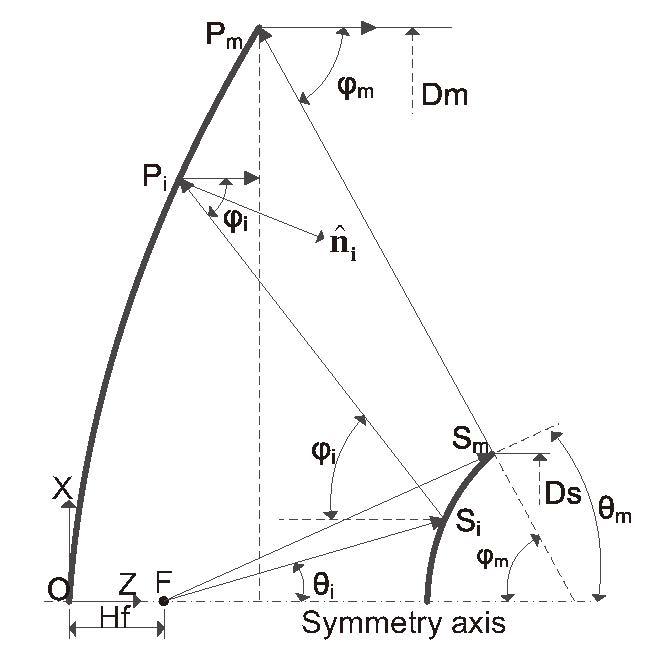}
	\caption{Schematic diagram of Description of Shaped Cassegrain dual-reflector. }
	\label{Fig1}
\end{figure}

%\noindent
In order to enable MEFCM to be applied in the analysis of structural deformation influence of shaped reflectors, we present a method for the description of the shaped reflector surface in this paper, which is based on standard discrete parabola set. The method mainly deals with electro-mechanical coupling analysis of shaped dual-reflector antennas. In order to obtain high aperture efficiency and low sidelobe levels, reflector shaping design is adopted to obtain the desired aperture field distribution. The shaping design should satisfy three conditions (\citealt{Milligan+2005}): conservation of power, equality of path-length and law of reflection. Shaped reflectors can spread spherical waves into a desired pattern based on geometric optics. The schematic diagram of the shaped Cassegrain dual-reflector antenna is shown in Figure \ref{Fig1}, where a spherical wave from the focal point F is changed into equal phase plane wave with the reflection by primary reflector and subreflector. With the shaping design, generatrices of the primary and secondary reflector are not standard parabola and hyperbola, respectively, and cannot be accurately descried by a closed form equation. The shaped surfaces do not satisfy the geometric relationship of classical Cassegrain systems, and thus, the classical dual-reflector OPD relationships (\citealt{Ruze+1969}) cannot be directly employed. In order to apply the Ruze’s OPD relationships, we described the shaped dual-reflector surface based on the standard discrete parabola set in this paper.

\subsection{Formulation Expression of Shaped Dual-Reflector}

\begin{figure}[htb]
	\centering
	\includegraphics[width=8cm, angle=0]{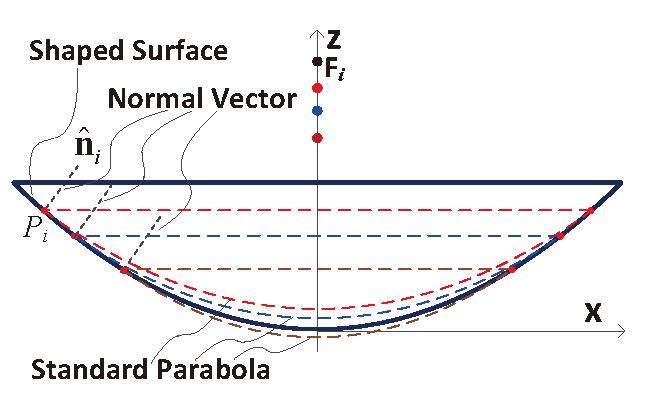}
	\caption{Description of shaped surface by standard discrete parabola set. }
	\label{Fig2}
\end{figure}

%\noindent
The diagram of the proposed description method for the dual-reflector surface is shown in Figure \ref{Fig2}, and the research object in the paper is the shaped Cassegrain antenna. The point set of the surface generatrix of shaped primary reflector is expressed as a discrete set of points of different parabolic equations, which is expressed as follow:
\begin{equation}
x_{i}^2=4f_{i}(z_{i}-h_{i})
\label{eq1}
\end{equation}

\noindent
where $h_{i}$  signifies the offset of Z direction of parabolic $i$ and $f_{i}$ signifies focal length of parabolic $i$. The point set of shaped dual-reflector are described as $L=\left \{ P_{1},\cdots, P_{n} \right \}$ for primary surface and $Q=\left \{ S_{1},\cdots, S_{n} \right \}$ for secondary surface, respectively, and there is a one-to-one correspondence between the points $P_{i}(x_{pi},z_{pi})$ and $S_{i}(x_{si},z_{si})$. The slopes of points on shaped primary surface, which are expressed as $k_{i}$, can be obtained by discrete point difference or mean weighted by areas of adjacent triangles. According to the characteristics of the parabola, $k_{i}$ is signified by:
\begin{equation}
k_{i}=x_{i}/(2f_{i})=\tan(\varphi _{i}/2)
\label{eq2}
\end{equation}

\begin{table}
	\begin{center}
		\caption[]{ Relative parameters of discrete normal parabola set.}\label{tab1}
 \begin{tabular}{clcl}
	\hline\noalign{\smallskip}
	Parameter &  Expression      & Parameter &  Expression                    \\
	\hline\noalign{\smallskip}
	Main-focal length (m)  & $f_{i}=x_{i}/(2\tan(\frac{\varphi _{i}}{2}))$ & Offset (m)     & $h_{i}=z_{i}-\frac{x_{i}^{2}}{4f_{i}}$  \\
	Magnification  & $M_{i}=\tan(\frac{\varphi _{i}}{2})/ \tan(\frac{\theta_{i}}{2})$     &   Eccentricity     & $e_{i}=\frac{M_{i}+1}{M_{i}-1}$\\
	Sub-focal length (m)  & $2c_{i}=\frac{x_{si}}{\tan(\varphi_{i})}+\frac{x_{si}}{\tan(\theta_{i})}$     &   Semi-major axis distance (m)     & $a_{i}=\frac{c_{i}}{e_{i}}$\\
	\noalign{\smallskip}\hline
\end{tabular}
\end{center}
\end{table}

As shown in Table \ref{tab1}, relative parameters of shaped dual-reflector can be obtained, where $\varphi _{i}$  and $\theta_{i}$  are the full angle between the incident and reflected rays and the angle between symmetry axis and ray from the subreflector to the feed, respectively.

\subsection{Formulation Expression of Optical Path Difference}

%\noindent
This paper is focused on the influence of antenna structure distortion on OPD and far field beam pattern. Because the errors caused by primary reflector surface distortion and subreflector displacement are generally not large, they belong to the small deformation errors compared with aperture size and focal length. Furthermore, as the displacement errors of reflector surface are always very small, the influence of the errors on the amplitude distribution of the aperture surface can be ignored, and only the influences on the phase distribution need to be considered.
%\noindent

The $\delta _{p}$, $\delta _{s}$ and $\delta _{f}$ represent OPDs caused by the deformation of the primary reflector, the offset of subreflector and the displacements of the feed, respectively, and the $\delta$ represents the sum of them. The OPD $\delta _{p}$ due to deformation of the primary reflector is expressed as follow:
\begin{equation}
\delta _{p}(r_{i},\phi  _{i})=\textup{\textbf{c}}_{\textup{pi}}\cdot \textup{\textbf{u}}_{\textup{pi}}
\label{eq3}
\end{equation}
where $\textup{\textbf{u}}_{\textup{pi}}=(\Delta x_{pi},\Delta y_{pi},\Delta z_{pi})$ is the displacement vector of primary surface. The components of $\textbf{c}_{\textup{pi}}$ are $c_{pi1}=-2n_{xi}n_{zi}$, $c_{pi2}=-2n_{yi}n_{zi}$, $c_{pi3}=-2n_{zi}^2$, and $n_{xi},n_{yi},n_{zi}$  signify the components of the unit normal vector. The OPD $\delta _{f}$ of the displacements of the feed is expressed as follow:
\begin{equation}
\delta _{f}(r_{i},\phi  _{i})=\textup{\textbf{c}}_{\textup{fi}}\cdot \textup{\textbf{u}}_{\textup{fi}}
\label{eq4}
\end{equation}
where $\textup{\textbf{u}}_{\textup{f}}=(\Delta x_{f},\Delta y_{f},\Delta z_{f})$ is the displacement vector of the feed. The components of $\textbf{c}_{\textup{fi}}$ are $c_{fi1}=-\sin\theta _{i}\cos\phi _{i}$, $c_{fi2}=-\sin\theta _{i}\sin\phi _{i}$, $c_{fi3}=1-\cos\theta _{i}$ . The OPD $\delta _{s}$ of the offset of the subreflector is expressed as follow:
\begin{equation}
\delta _{s}(r_{i},\phi  _{i})=\textup{\textbf{c}}_{\textup{si}}\cdot \textup{\textbf{p}}
\label{eq5}
\end{equation}
where $\textup{\textbf{p}}=[\Delta x_{s},\Delta y_{s},\Delta z_{s},\Delta \gamma _{x},\Delta \gamma _{y}]^{^{T}}$ is the offset vector of the subreflector. The components of $\textbf{c}_{\textup{si}}$  are  $c_{si1}=(\sin\theta _{i}-\sin\varphi _{i})\cos\phi _{i}$, $c_{si2}=-(\sin\theta _{i}-\sin\varphi _{i})\sin\phi _{i}$, $c_{si3}=-(\cos\varphi _{i}+\cos\phi _{i})$,  $c_{si4}=-(c_{i}-a_{i})(\sin\theta _{i}+M_{i}\sin\varphi _{i})\sin\phi _{i}$, $c_{si5}=-(c_{i}-a_{i})(\sin\theta _{i}+M_{i}\sin\varphi _{i})\cos\phi _{i}$.

%=======================Section 3===============================
\section{Mechanical-electromagnetic-field-coupling-model}
\label{sect:Set3}

\begin{figure}[htb]
	\centering
	\includegraphics[width=8cm, angle=0]{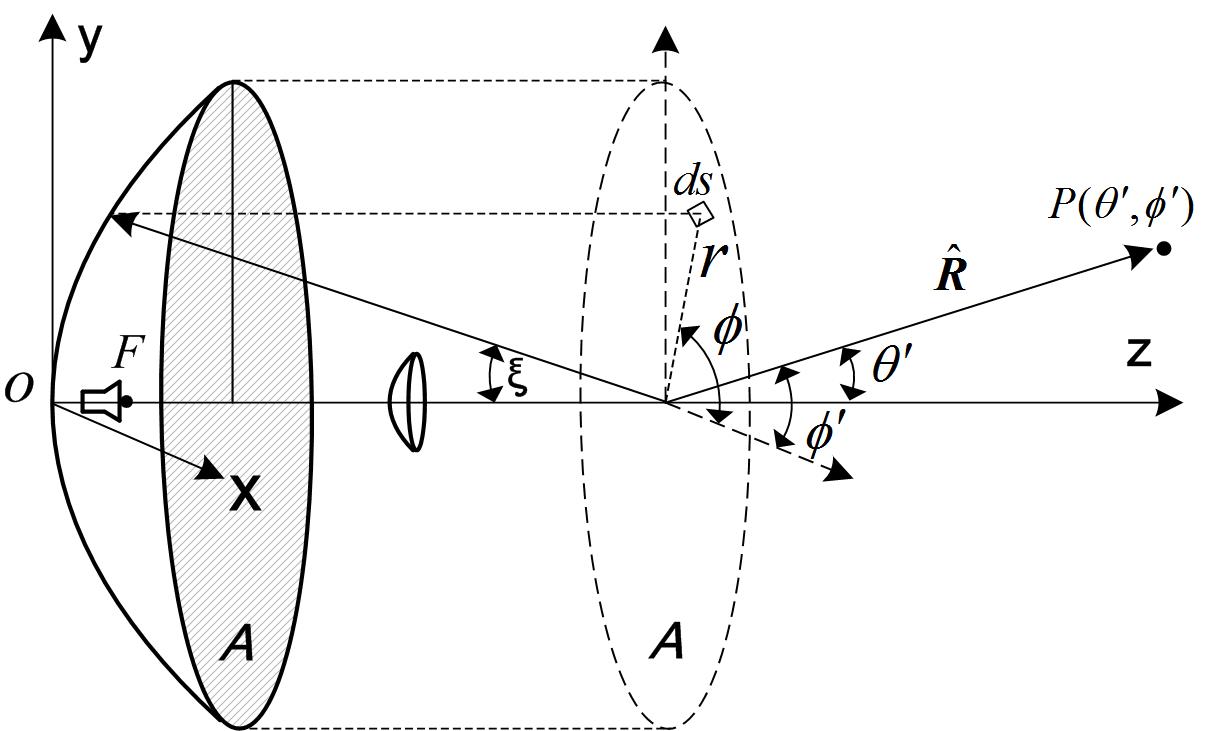}
	\caption{Geometric diagram dual-reflector antenna.}
	\label{Fig3}
\end{figure}

In this paper, the aperture field method is adopted to calculate the far field radiation pattern of reflector antennas. The method could obtain the aperture field distribution from radiation field of the feed by geometrical optics. According to the Fourier transform relationship between aperture field and far field, the MEFCM of standard reflector (Figure \ref{Fig3}) is expressed as follow:
\begin{equation}
	T_{1}(\theta^\prime,\phi^\prime)=\iint_{A}F(r,\phi)\exp[jkr\sin{\theta^\prime}\cos{(\phi^\prime-\phi)}]\cdot\exp (j\Delta\varphi_1)rdrd\phi
	\label{eq6}
\end{equation}
and expressed the sum of discrete segments for shaped surface as follow:
\begin{equation}
	T_{1}(\theta^\prime,\phi^\prime)=\sum_{i=1}^{m}F(r_{i},\phi_{i})\exp[jkr_{i}\sin{\theta^\prime}\cos{(\phi^\prime-\phi_{i})}]\cdot \exp(jk\delta _{i})\Delta s_{i}
	\label{eq7}
\end{equation}
\noindent
where $F(r,\phi)$ is the aperture field distribution function of standard reflector on the aperture surface $A$ ; $(\theta^\prime,\phi^\prime)$ is coordinates of the observation direction in far field and $(r,\phi)$ is coordinates of the point in aperture plane; $k$ is the free space wave constant, $k=2\pi/\lambda$, $\lambda$ is the working wavelength. $\Delta \varphi_{1}$ is phase error distribution function in aperture plane, and $\Delta \varphi_{1} =k\cdot \delta$, $\delta$ is the total aperture OPD. The Gauss Integration method can be adopted to solve the MEFCM.

%=======================Section 4===============================
\section{Determination of Subreflector adjustment position}
\label{sect:Set4}
%%---------4.1-----------
\subsection{Additional error beams}
Because the antenna structure deformation was only small displacement, low order Taylor series expansion is adopted and the Eq.\ref{eq6} can be derived as follow:
\begin{equation}
	\begin{split}
	T_{1}\left(\theta^\prime,\phi^\prime\right)=&\iint_{A}F\left(r,\phi\right)\exp\left[jkr\sin{\theta^\prime}\cos{\left(\phi^\prime-\phi\right)}\right]\cdot\left(1+j\Delta\varphi_1\right)rdrd\phi \\
	&=T_{0}\left(\theta^\prime,\phi^\prime\right)+T^\prime\left(\theta^\prime,\phi^\prime\right)
	\end{split}
	\label{eq8}	
\end{equation}
\noindent
where $T_{0}$ signifies the ideal beam pattern and $T^\prime$ signifies  the error beam pattern, and the equations could be expressed as follow:
\begin{equation}
	T_{0}(\theta^\prime,\phi^\prime)=\iint_{A}F(r,\phi)\exp[jkr\sin{\theta^\prime}\cos{(\phi^\prime-\phi)}]rdrd\phi
	\label{eq9}
\end{equation}

\begin{equation}
	T^\prime(\theta^\prime,\phi^\prime)=\iint_{A}F(r,\phi)\exp[jkr\sin{\theta^\prime}\cos{(\phi^\prime-\phi)}]\cdot(j\Delta\varphi_1)rdrd\phi
	\label{eq10}
\end{equation}

According to Eq. \ref{eq8}, the beam pattern $T_{1}$ caused by structure deformation could be understood as the synthesis of far field ideal beam $T_{0}$ and error beam $T^\prime$. For the practical antennas, the ideal beam can be calculated by radiation integral of designed aperture distribution function and expressed as $\hat{T}_0$; $T_1$ can be obtained by total power measurement of the strong radiation source, such as the artificial satellite, and expressed as $\widetilde{T}_1$. Then, Eq. \ref{eq8} can be written as:
\begin{equation}
	\widetilde{T}_1(\theta^\prime,\phi^\prime)=\hat{T}_0(\theta^\prime,\phi^\prime)+T^\prime(\theta^\prime,\phi^\prime)
	\label{eq11}
\end{equation}

The error beam can be written as:
\begin{equation}
	T^\prime(\theta^\prime,\phi^\prime)=\widetilde{T}_1(\theta^\prime,\phi^\prime)-\hat{T_0}(\theta^\prime,\phi^\prime)
	\label{eq12}
\end{equation}

The subreflector adjustment will cause additional OPD and additional error beam. After the subreflector adjustment, the far field beam pattern can be written as:
\begin{equation}
	T_2(\theta^\prime,\phi^\prime)={\widetilde{T}}_1(\theta^\prime,\phi^\prime)+{\hat{T}}^{\prime\prime}(\theta^\prime,\phi^\prime)
	\label{eq13}
\end{equation}

Let the parameter vector of subreflector adjustment be expressed as $\textbf{p}$. Then, the additional error beam is written as ${\hat{T}}^{\prime\prime}(\theta^\prime,\phi^\prime,\textbf{p})$, and can be calculated by Eq. \ref{eq10}. According to Eq. \ref{eq5}, the additional phase error is expressed as：

\begin{equation}
	\mathrm{\Delta}\varphi_1=g\left(\textbf{p}\right)=k\cdot\textbf{c}_{si}\cdot\textbf{p}
	\label{eq14}
\end{equation}

%%---------4.2-----------
\subsection{Iteration approach of subreflector adjustment parameters}
It is obvious that the electromagnetic performance of the perfect reflector is the optimal one, that is to say, the gain is the maximum and the beam is symmetric, as ideal beam pattern. For Eq. \ref{eq13}, both sides are subtracted by $\hat{T}_0$, resulting in:

\begin{equation}
	T_2(\theta^\prime,\phi^\prime)-{\hat{T}}_0(\theta^\prime,\phi^\prime)={\widetilde{T}}_1(\theta^\prime,\phi^\prime)-\hat{T}_0(\theta^\prime,\phi^\prime)+{\hat{T}}^{\prime\prime}(\theta^\prime,\phi^\prime,\textbf{p})
	\label{eq15}
\end{equation}
and let beam deviation be
\begin{equation}
	L={\widetilde{T}}_1(\theta^\prime,\phi^\prime)-{\hat{T}}_0(\theta^\prime,\phi^\prime)+{\hat{T}}^{\prime\prime}(\theta^\prime,\phi^\prime,\textbf{p})
	\label{eq16}
\end{equation}

Note that the Eq. \ref{eq15} can be understood as deviation between beam pattern after subreflector adjustment and ideal beam pattern.  As a consequence, it is considered that the numerical iterative techniques can be adopted to determine the optimal parameter of subreflector position. Thus, the parameter iteration method can be adopted to improve the electromagnetic performance, and minimize the deviation. The deviation can be expressed as:

\begin{equation}
	V_{l}=\|{\widetilde{T}}_1(\theta^\prime,\phi^\prime)-{\hat{T}}_0(\theta^\prime,\phi^\prime)+{\hat{T}}^{\prime\prime}(\theta^\prime,\phi^\prime,\textbf{p})\|
	\label{eq17}
\end{equation}
where $\|\cdot \|$  is the operator of  pattern vector. Generally, the infinity norm can be used.

\begin{figure}[h]
	\centering
	\includegraphics[width=8cm, angle=0]{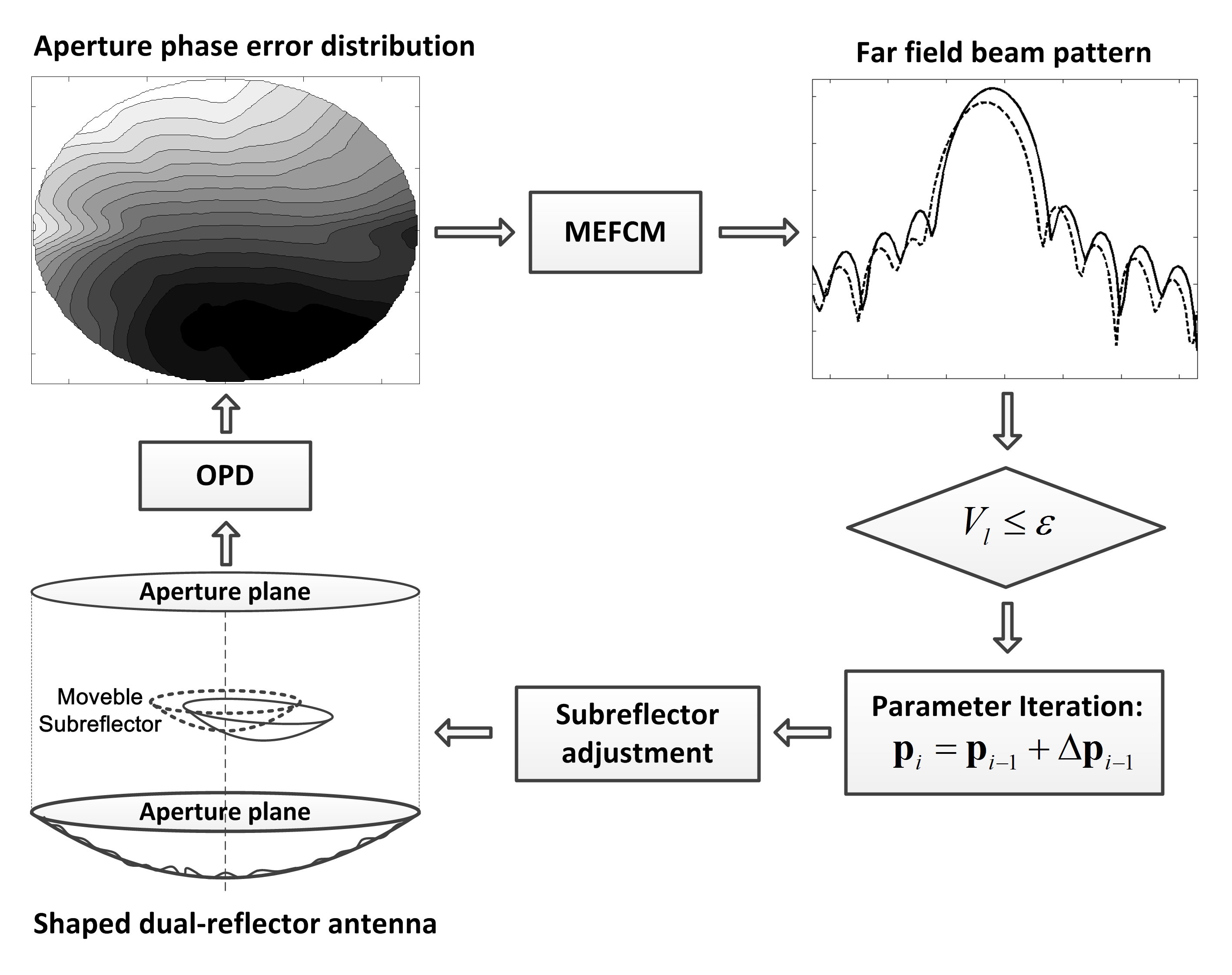}
	\caption{Schematic diagram of antenna performance improvement based on subreflector adjustment parameters iteration.}
	\label{Fig4}
\end{figure}

When the antenna points at a radio source at an elevation angle, the backup structure will be distorted due to the gravity, and then, surface errors of primary reflector and displacement of subreflector will be produced, which will lead to the degradation of the EM performance of the antenna. The schematic diagram of antenna performance improvement based on subreflector adjustment parameters iteration is shown in Figure \ref{Fig4}. By repeating the adjustment procedure and iteration process again and again, the beam pattern could be gradually improved.

\begin{figure}[h]
	\centering
	\includegraphics[width=8cm, angle=0]{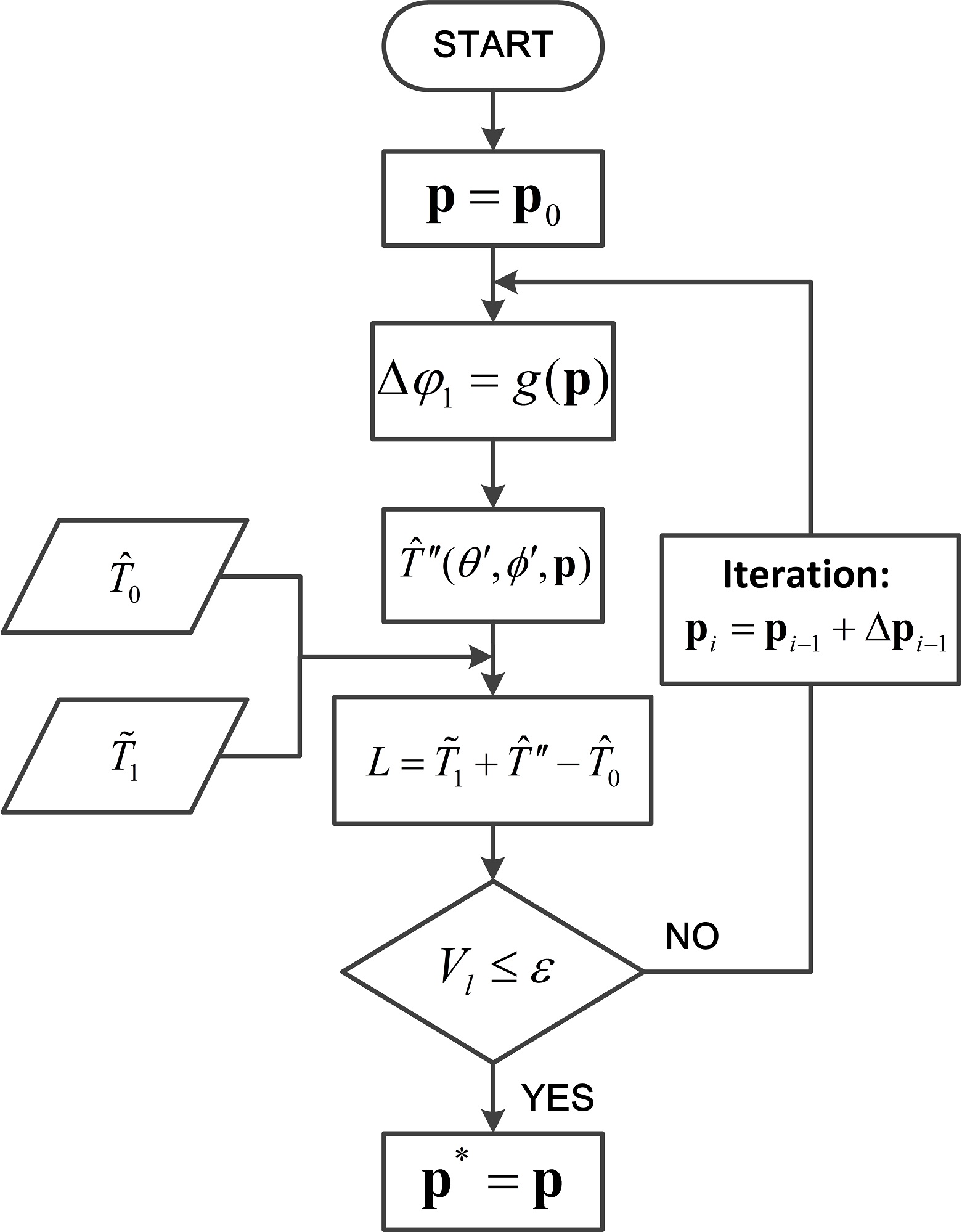}
	\caption{Flow chart of iteration procedure.}
	\label{Fig5}
\end{figure}
Performance improvement process based on subreflector adjustment and parameters iteration is shown in Figure \ref{Fig5}, and can be described as follow:
\begin{enumerate}[(1) ]
	\item Set the iterative initial value as $\textbf{p}_0$;  %(1)
	\item According to Eq. \ref{eq14},  the additional OPD and phase error $\Delta\varphi_1$ can be calculated ;	%(2)
	\item According to Eq. \ref{eq10},  the additional error beam ${\hat{T}}^{\prime\prime}(\theta^\prime,\phi^\prime,\textbf{p})$ can be calculated ; 	%(3)
	\item According to Eq. \ref{eq16},  the beam deviation after subreflector adjustment can be obtained;		%(4)
	\item According to Eq. \ref{eq17}, the iterative objective function $L$ can be calculated and the judgment will be made according to the expression $V_{l}\leqslant \varepsilon$, where $\varepsilon$ is the relatively altering boundary. Through a large number of simulations, we think that the $\varepsilon$ may be generally less than or equal to 1 dB.\\
	If the expression cannot be satisfied, the subreflector adjustment parameter vector $\textbf{p}$ will be modified by $\textbf{p}_i=\textbf{p}_{i-1}+\Delta \textbf{p}_{i-1}$, and the process returns to step (2).\\
	If the expression can be satisfied, the iteration process terminates, and the current parameter $\textbf{p}$ is the optimal parameter $\textbf{p}^{*}$.		%(5)
\end{enumerate}

%=======================Section 5===============================
\section{Example and discussion}
\label{sect:Set5}

\begin{figure}[h]
	\centering
	\includegraphics[width=8cm, angle=0]{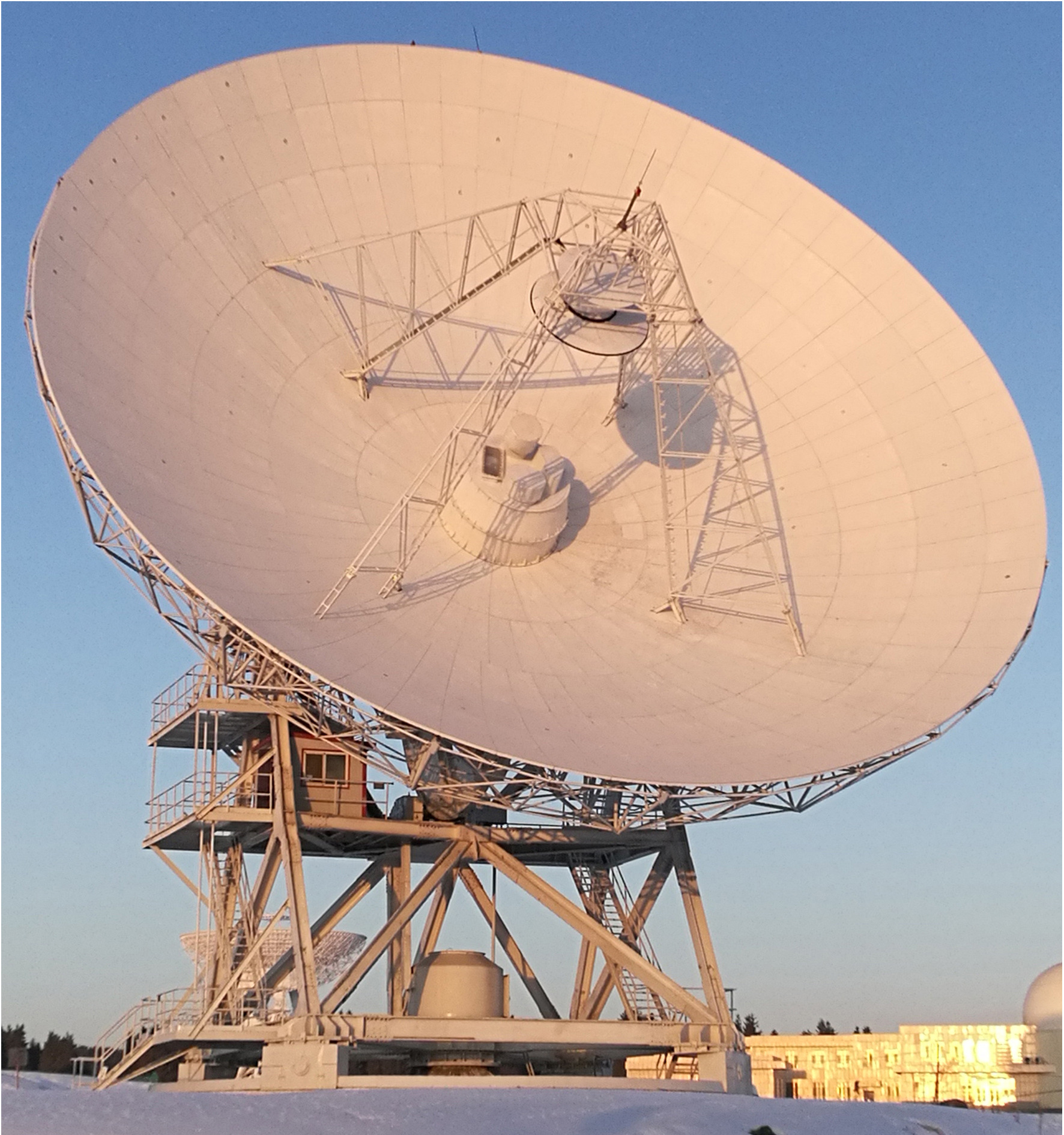}
	\caption{The image of a 25 m shaped Cassegrain antenna.}
	\label{Fig6}
\end{figure}

\begin{figure}[h]
	\centering
	\includegraphics[width=8cm, angle=0]{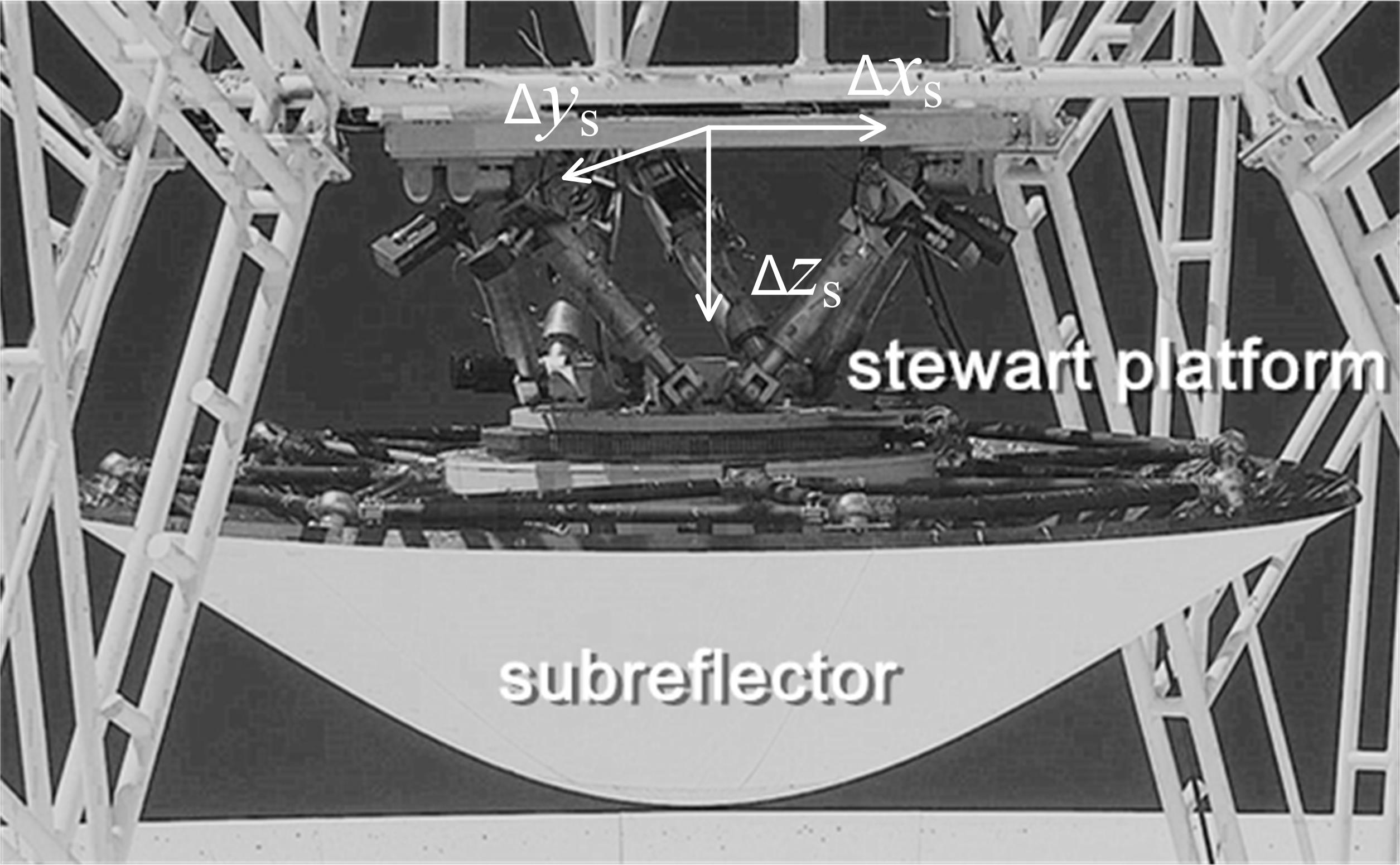}
	\caption{Subreflector equipped with Stewart platform.}
	\label{Fig7}
\end{figure}

\begin{figure}[htb]
	\centering
	\includegraphics[width=8cm, angle=0]{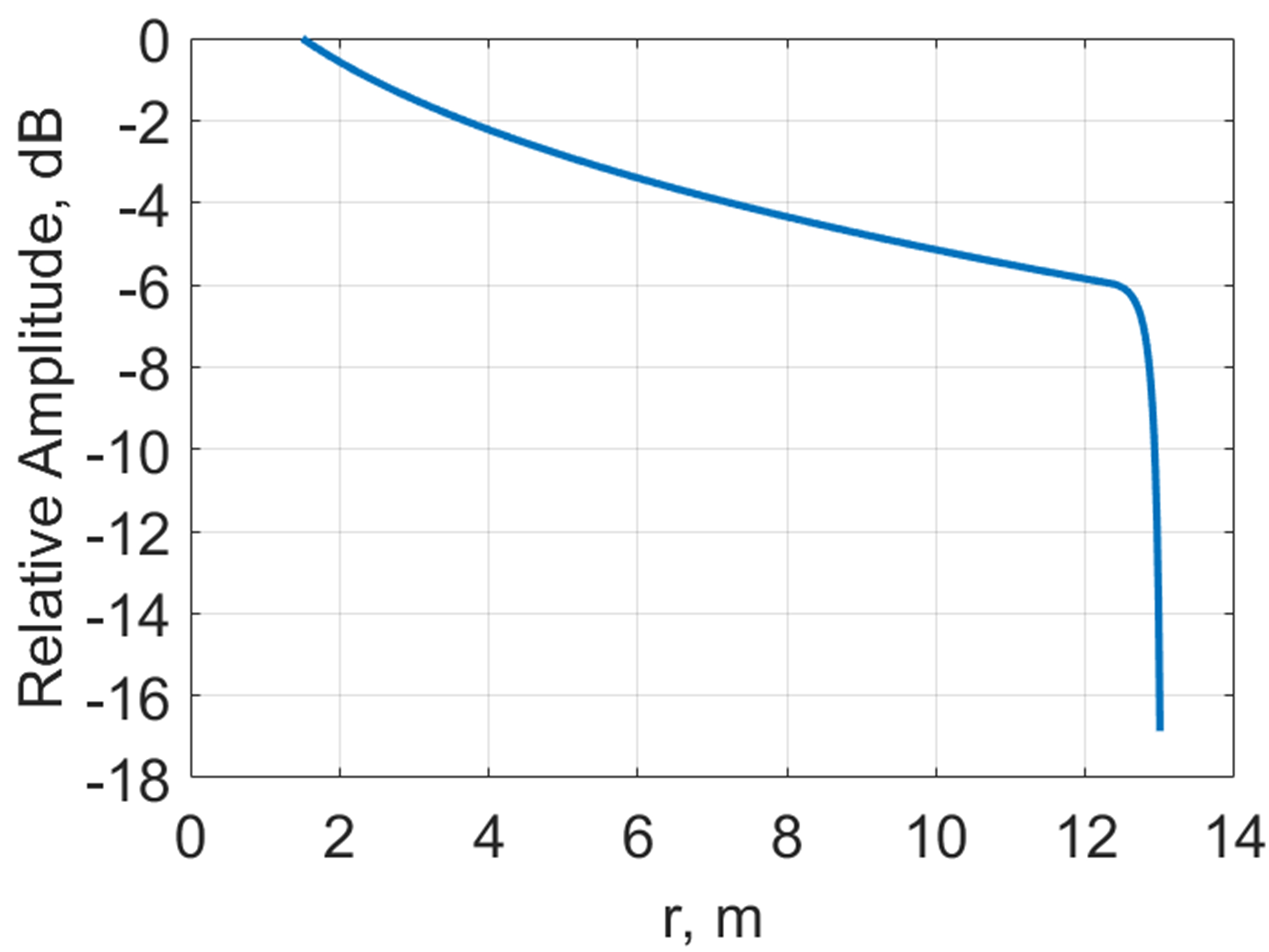}
	\caption{Aperture field distribution.}
	\label{Fig8}
\end{figure}

The example of a 25 m shaped Cassagrain antenna has been analysed. In the example, the antenna is used for the radio astronomical observation with C band as the main operating wave band (wavelength 6 cm, frequency 4.8 GHz). The 25 m shaped Cassagrain antenna is shown in Figure \ref{Fig6}, and a 6 DoF (Degree of Freedom) Stewart platform has been installed in the quadripod to adjust subreflector position, which is shown in Figure \ref{Fig7}. The 5 degrees of freedom of Stewart platform was used primarily to adjust subreflector in real-time and the adjustment ranges are given in Table \ref{tab2}. The basic normal parameters of 25 m Shaped cassegain antenna are given in Table \ref{tab3} and the aperture field distribution is shown in Figure \ref{Fig8}, which will be substituted into the MEFCM as the $F(r,\phi)$.

\begin{table}
	\bc
	\begin{minipage}[]{100mm}
		\caption[]{Subreflector adjustment ranges.}\label{tab2}\end{minipage}
	\setlength{\tabcolsep}{2.5pt}
	\small
	\begin{tabular}{ccccccccccccc}
		\hline\noalign{\smallskip}
		Parameter &  $\Delta x_s$ (mm) & $\Delta y_s$ (mm) & $\Delta z_s$ (mm) & $\Delta \gamma _x (^{\circ})$ &
		$\Delta \gamma _y (^{\circ})$ \\
		\hline\noalign{\smallskip}
		Range  & $\leqslant 50$ & $\leqslant 50$ & $\leqslant 80$ & $\leqslant 5$ & $\leqslant 5$ \\
		\noalign{\smallskip}\hline
	\end{tabular}
	\ec
\end{table}

\begin{table}
	\bc
	\begin{minipage}[]{120mm}
		\caption[]{Basic nominal parameters of the 25 m Shaped Cassegrain antenna.}\label{tab3}\end{minipage}
	\setlength{\tabcolsep}{2.5pt}
	\small
	\begin{tabular}{ccccccccccccc}
		\hline\noalign{\smallskip}
		Parameter & value & Parameter & value \\
		\hline\noalign{\smallskip}
		Main reflector diameter (m)	&25	&Subreflector diameter (m)	&3 \\
		Half-angle subtended of main reflector ($^{\circ}$)	&79.61	&Half-angle subtended of subreflector ($^{\circ}$)	&14.43 \\
		Main reflector focal length (m)	&7.8  &	Hyperboloid focal length (m)	&0.530\\
		Eccentricity	&1.358	&Magnification	&6.583 \\
		\noalign{\smallskip}\hline
	\end{tabular}
	\ec
\end{table}

%%---------5.1-----------
\subsection{Shaped surface and structural analysis}

\par
\begin{figure}[htb]
	\centering
	\includegraphics[width=8cm, angle=0]{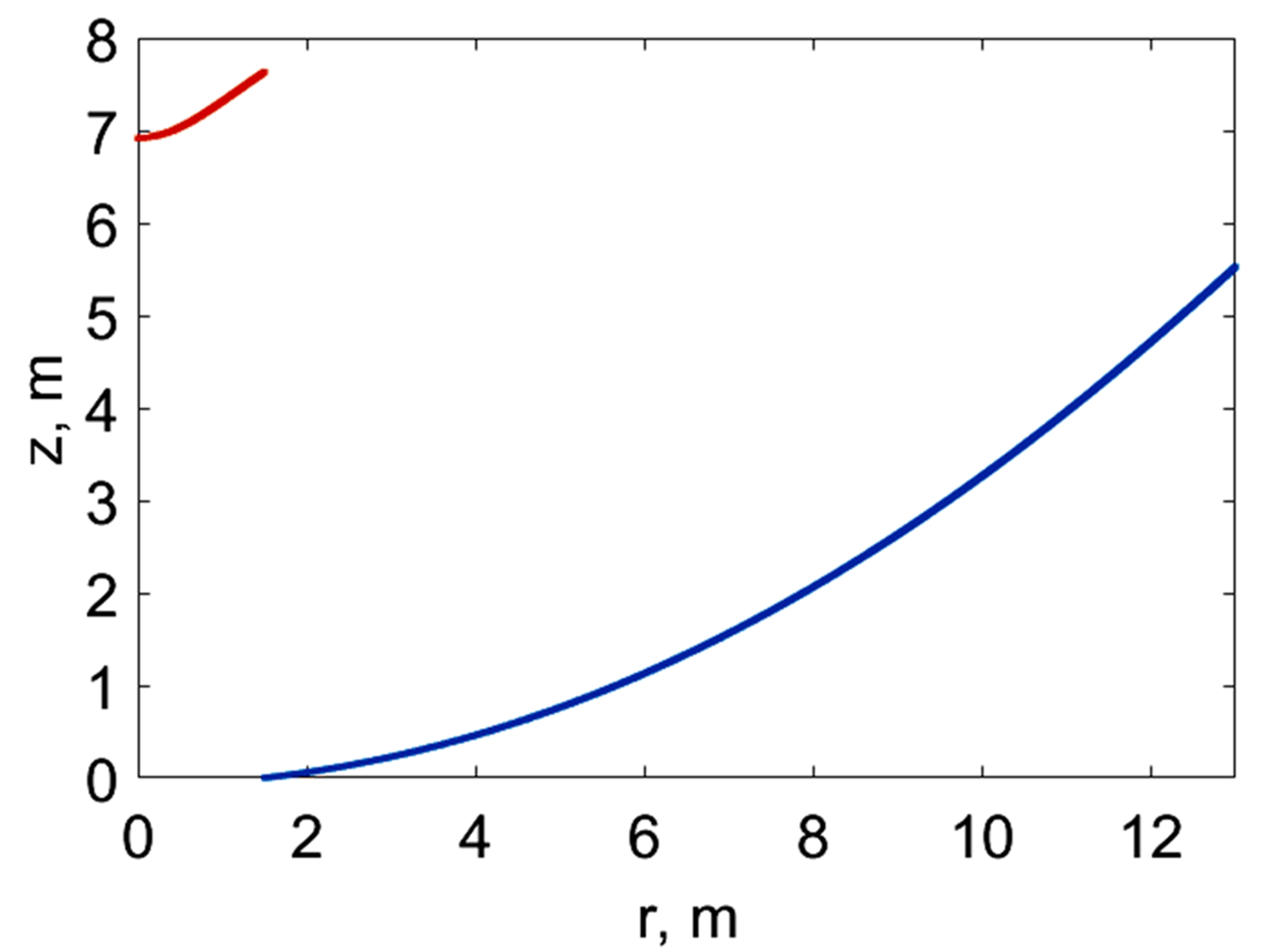}
	\caption{Shaped Cassegrain antenna generatrix.}
	\label{Fig9}
\end{figure}

\setlength\parindent{2em}The shaped surface of the shaped Cassegrain antenna is described by discrete standard parabola set and the generatrices of the main reflector and subreflector are both made up of 7604 discrete points, which are obtained by Eq. \ref{eq1}- \ref{eq2} and shown in Figure \ref{Fig9}. Relative parameters of discrete normal parabola set are computed by expressions given in Table \ref{tab1}, and OPD and far field beam pattern can be computed by Eq. \ref{eq3}- \ref{eq5} and Eq. \ref{eq7}.

\begin{figure}[htb]
	\centering
	\includegraphics[width=8cm, angle=0]{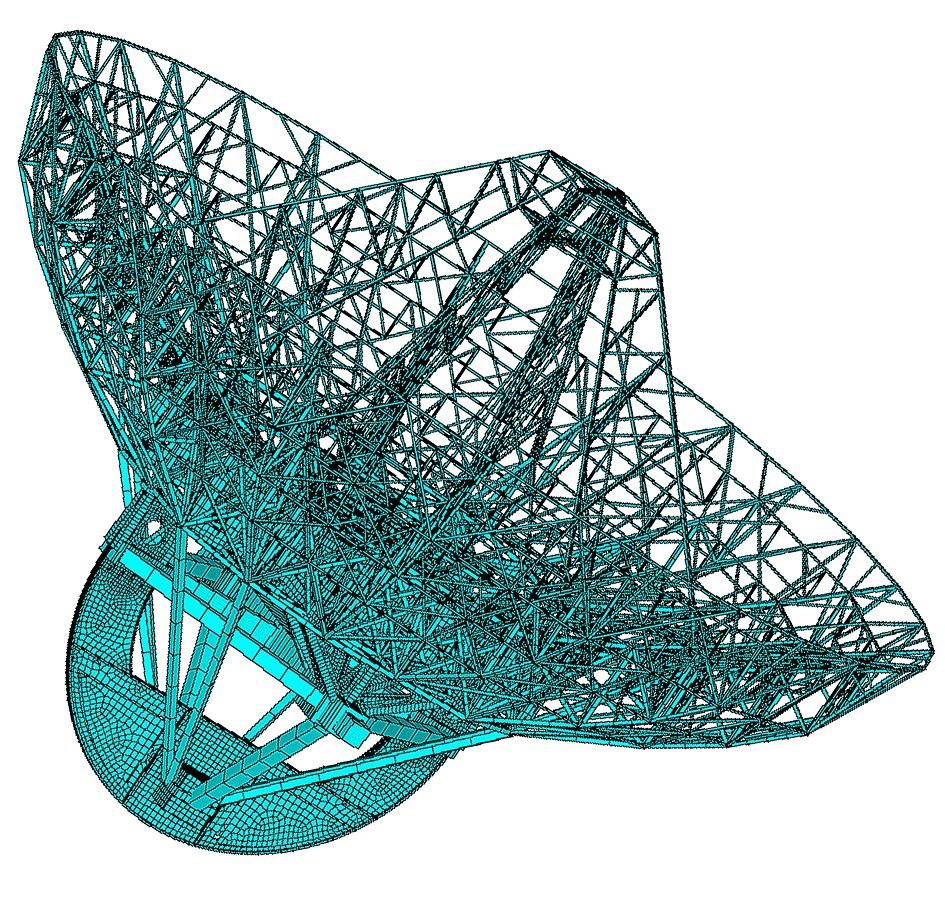}
	\caption{Antenna reflector FE model.}
	\label{Fig10}
\end{figure}

\par To obtain antennas far field beam due to structural deformation, a Finite Element (FE) model of antenna reflector structure was established, which was shown in Figure \ref{Fig10}, and gravitational-structural deformations at 40° and 70° elevation angles were analyzed. The node displacements of backup structure and quadripod structure were obtained and then, reflector surface distortion and subreflector rigid displacement were computed by interpolation and coordinate transformation.

%%---------5.2-----------
\subsection{Parameter iterative results}

\par
\begin{figure}[htb]
	\centering
	\includegraphics[width=14cm, angle=0]{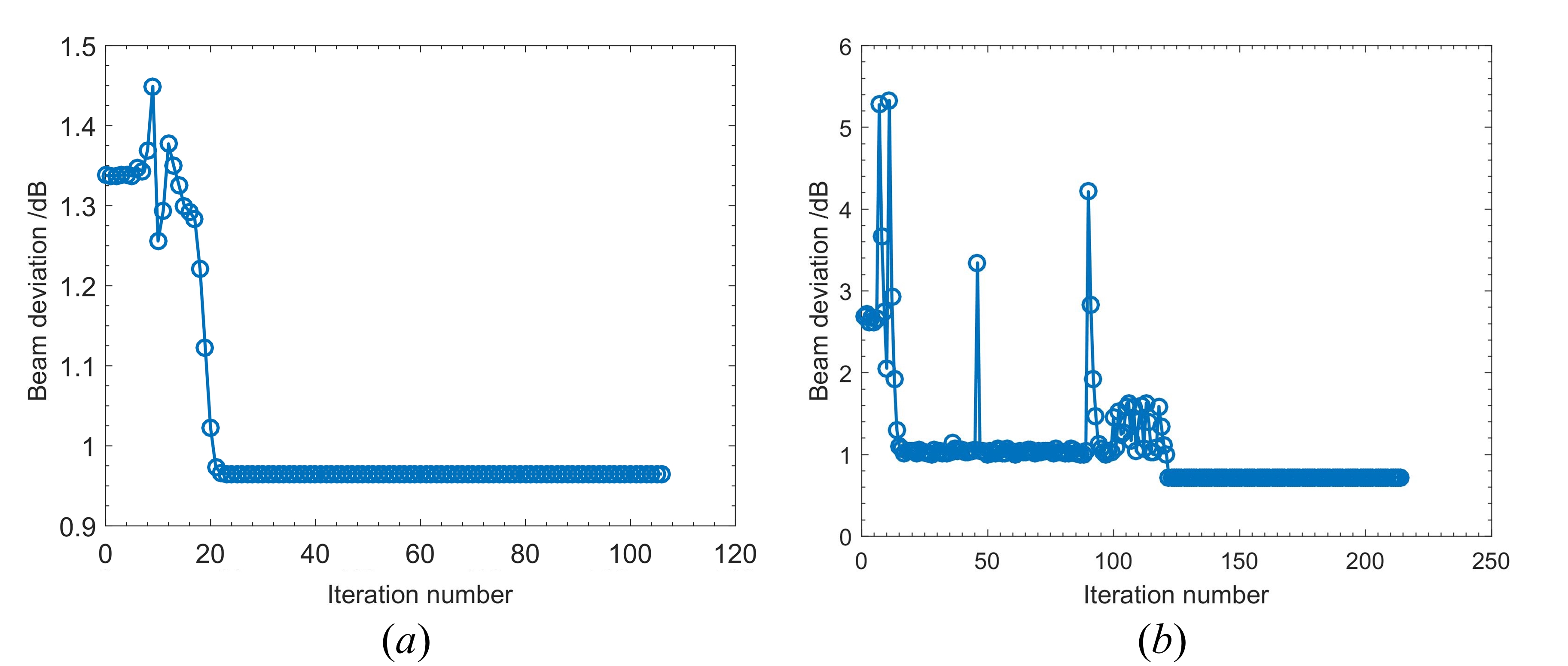}
	\caption{ Iteration processes for adjustment parameters.  ($a$) for 40° elevation angle, ($b$) for 70° elevation angle. }
	\label{Fig11}
\end{figure}

 A computer program of numerical computation has been generated on the basis of the theoretical developments presented above to achieve parameters iteration and beam pattern calculation. To obtain the global optimum solution rapidly, a constraint condition is added to the iteration process, that is, the pointing error should be less than 0.1*HPBW (Half Power Beam Width), and the interior-point method is adopted in the iteration process. The numerical computation has been performed on basis of the FE model. The iteration processes in the cases of 40° and 70° elevation angles are shown in Figure \ref{Fig11}. In the cases of 40° and 70° elevation angles, the optimal values are obtained after 107 and 214 iterations, and the optimal parameters are $\textbf{p}^{*}=[0.5793, -0.2229, -9.9038, -0.0007, 0.0019]$, and $\textbf{p}^{*}=[0.1223, -5.6443, -3.0728, -0.0019, 0.0002]$, respectively.

\begin{figure}[h]
	\centering
	\includegraphics[width=14cm, angle=0]{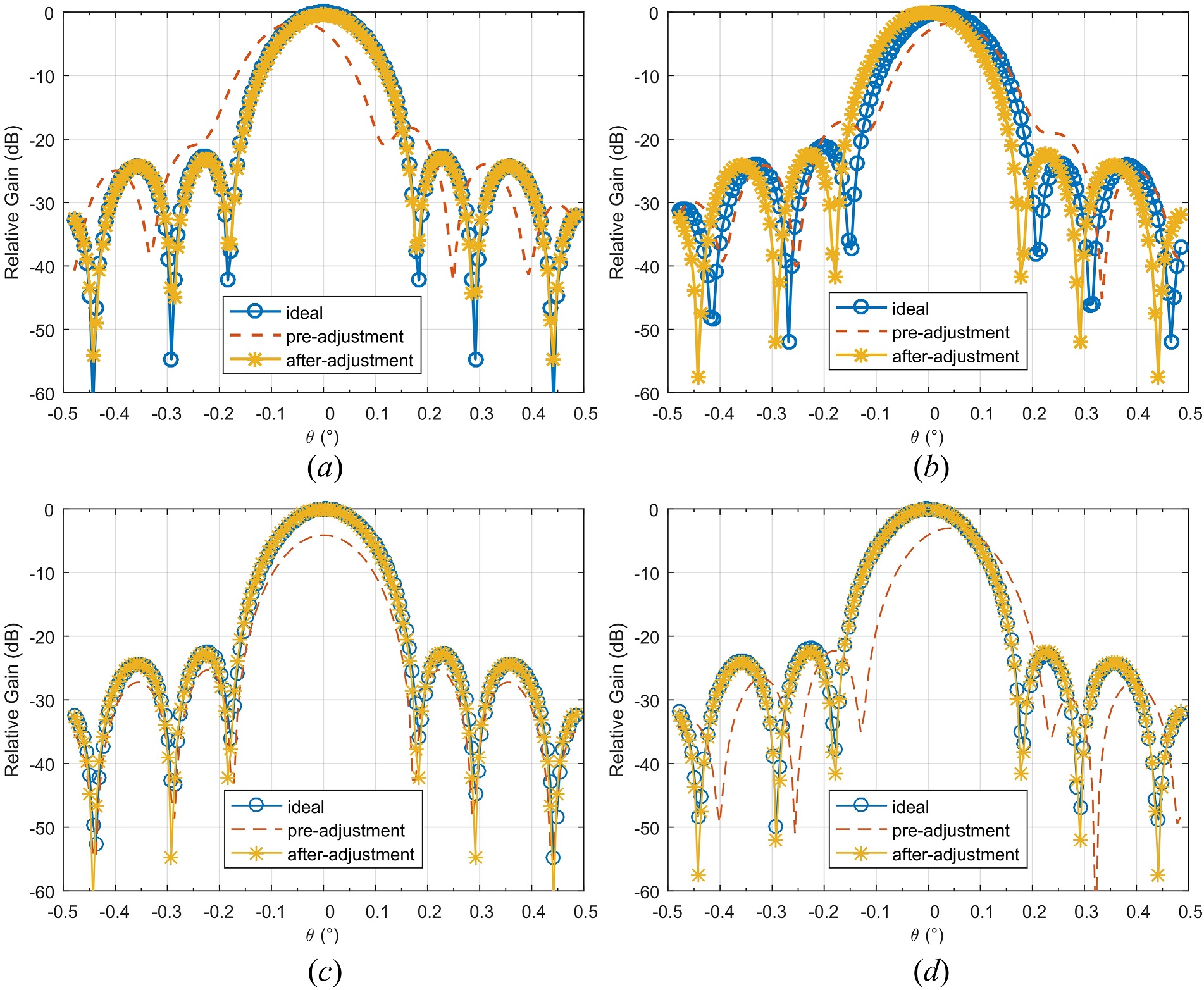}
	\caption{ Far field beam patterns for subreflector adjustment.  ($a$)  0$^{\circ}$ cut plane for 40° elevation angle, ($b$) 90$^{\circ}$ cut plane for 40° elevation angle; ($c$)  0$^{\circ}$ cut plane for 70° elevation angle, ($d$) 90$^{\circ}$ cut plane for 70° elevation angle.}
	\label{Fig12}
\end{figure}

\begin{figure}[h]
	\centering
	\includegraphics[width=14cm, angle=0]{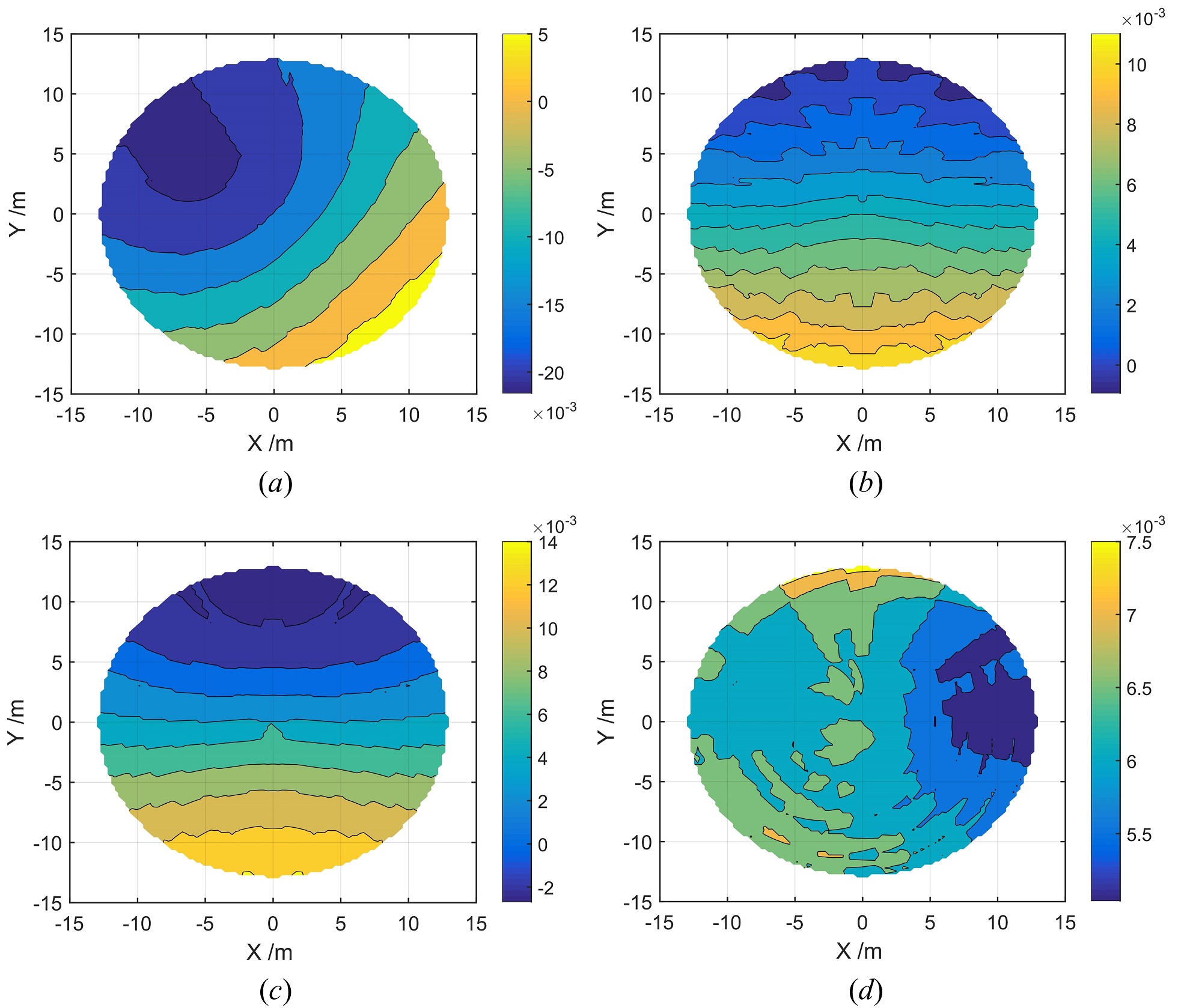}
	\caption{{\color{blue} OPD distributions in aperture plane for subreflector adjustment.  ($a$)  pre-adjustment for 40° elevation angle, ($b$)  after-adjustment for 40° elevation angle; ($c$)  pre-adjustment for 70° elevation angle, ($d$)  after-adjustment for 70° elevation angle.} }
	\label{Fig13}
\end{figure}

\par The effect of subreflector position adjustment based on optimal parameter $\textbf{p}^{*}$ is shown in Figure \ref{Fig12}. Figures show that after the subreflector adjustment the beam pattern is close to the ideal beam pattern.  Furthermore, from Figure \ref{Fig12} ($a$) and ($c$) as 0$^{\circ}$ cut plane (azimuth scan), the beam patterns after the adjustment almost coincide with the ideal beam pattern which indicate that reflector gravitational-structural deformations of azimuth direction have been fully compensated. In Figure \ref{Fig12} ($b$) and ($d$) as 90$^{\circ}$ cut plane (elevation scan), the shapes of beam pattern of after-adjustment approximates the ideal one, and they are just slightly different in beam center and sidelobes, which indicate that reflector gravitational-structural deformation of elevation direction have also been well compensated. The OPD distributions in aperture plane for subreflector adjustment are shown in Figure \ref{Fig13}.  It can be seen that the OPD after subreflector adjustment has been decreased significantly.

%\end{figure}

The results indicate that, for the 25 m Shaped Cassegrain antenna, the gravitational-structural deformations of the main reflector are almost homologous, that is, the main reflector almost is deformed into an approximate parabolic shape, a significant loss of beam pattern may occur as a result of subreflector position misalignment. Furthermore, the distortion of beam pattern of the 25 m Shaped Cassegrain antenna may be substantially compensated by suitable adjustments of the position of the subreflector and the parameter is effective and appropriate.

%=======================Section 6===============================
\section{Conclusion}
\label{sect:Set6}

\par A new method to determine the optimal subreflector position for the shaped Cassegrain antennas has been presented. The method is based on parameters iteration of subreflector adjustment to compensate effect of structural deformation of reflector. Considering the particularity of the shaped reflector antenna, we describe the shaped dual-reflector surface as the discrete normal parabola set accurately and utilize the features of classical Cassegrain system in the OPD relationship. By decomposing subreflector adjustment parameter with MEFCM, the iteration method of adjustment parameter can be adapted to improve the antenna EM performance, and the optimal adjustment parameter can be obtained rapidly by numerical computation.  An example of a 25 m Shaped Cassegrain antenna has been performed, and the results indicate that the methods proposed in this paper are effective and can be used in practice engineering.

%=======================Section 7===============================
%\normalem
%\begin{acknowledgements}
%
%
%
%
%\end{acknowledgements}

\normalem
\begin{acknowledgements}
	This work was supported by the National Key Basic Research Program of China under Grant [number 2015CB857100]; the Chinese Academy of Sciences (CAS) "Light of West China" Program under Grant [number 2017-XBQNXZ-B-021], and the National Natural Science Foundation of China under Grant [numbers 51522507, 51475349]. The work was also partly supported by the Operation, Maintenance and Upgrading Fund for Astronomical Telescopes and Facility Instruments, budgeted from the Ministry of Finance of China (MOF) and administrated by CAS.
	The authors thank the antenna engineers of the No. 39 Research Institute, CETC, for their valuable assistance.
	
\end{acknowledgements}

%\bibliographystyle{raa}
%\bibliographystyle{plain}
%\bibliography{xbb}

\end{document}